\documentclass{llncs}
\usepackage{amsmath,amsfonts,amssymb,latexsym,eucal}
\usepackage[colorlinks=true]{hyperref}
\usepackage{microtype}
\usepackage{pdfsync}
\usepackage[numbers]{natbib}

\newcommand{\remove}[1]{}

\newcommand{\BBB}{\vspace*{-\bigskipamount}}

\newtheorem{fact}{Fact}

\newcommand{\mc}{\mathcal}

\DeclareMathOperator*{\Exp}{\mathbb{E}}
\newcommand{\polylog}{\operatorname{polylog}}

\DeclareMathOperator{\Ext}{Ext}
\DeclareMathOperator{\rejsam}{\textsc{Rej}}

\title{Randomness Efficient Steganography}
\author{}
\institute{}
\date{}
\begin{document}

\maketitle

\begin{abstract}
  Steganographic protocols enable one to embed covert messages into
  inconspicuous data over a public communication channel in such a way
  that no one, aside from the sender and the intended receiver, can
  even detect the presence of the secret message. In this paper, we
  provide a new provably-secure, private-key steganographic encryption
  protocol secure in the framework of Hopper et
  al~\cite{DBLP:conf/crypto/HopperLA02}.  We first present a
  ``one-time stegosystem'' that allows two parties to transmit
  messages of fixed length (depending on the length of the shared key)
  with \emph{information-theoretic} security guarantees. Employing a
  pseudorandom generator (PRG) permits secure transmission of longer
  messages in the same way that such a generator allows the use of
  one-time pad encryption for long messages in a symmetric encryption
  framework. The advantage of our construction, compared to all
  previous work is \emph{randomness efficiency}: in the information
  theoretic setting our protocol embeds a message of length $n$ bits
  using a shared secret key of length $(1+o(1))n$ bits while achieving
  security $2^{-n/\polylog n}$; simply put this gives a rate of key
  over message that is 1 as $n\rightarrow\infty$ (the previous best
  result~\cite{stego1} achieved a constant rate $>1$ regardless of the
  security offered). In this sense, our protocol is the first truly
  randomness efficient steganographic system and breaks through a
  natural barrier imposed by bounded-round rejecting
  sampling. Furthermore, in our protocol, we can permit a portion of
  the shared secret key to be \emph{public} while retaining precisely
  $n$ private key bits. In this setting, by separating the public and
  the private randomness of the shared key, we achieve security of
  $2^{-n}$. Our result comes as an effect of a novel application of
  randomness extractors to stegosystem design.
% To the best of our knowledge this is the first time
 % extractors have been applied in steganography.
\end{abstract}

%\noindent
%\textbf{Keywords:} Information hiding, steganography, data hiding, steganalysis, covert communication.

\section{Introduction}

The steganographic communication problem can be described using
Simmons'~\cite{DBLP:conf/crypto/Simmons83} formulation of the problem:
Alice and Bob are prisoners who wish to communicate securely in the
presence of an adversary, called the ``Warden.'' The warden monitors
the communication channel to detect whether they exchange
``conspicuous'' messages. In particular, Alice and Bob are permitted
to exchange messages that adhere to certain channel distributions that
represent ``inconspicuous'' communication, but may not detectably
stray from this distribution. By controlling the messages transmitted
over such a channel, however, Alice and Bob may in fact exchange
messages that cannot be detected by the Warden.  There have been two
approaches in formalizing this problem, one based on information
theory~\cite{DBLP:conf/ih/Cachin98,DBLP:conf/ih/ZollnerFKPPWWW98,DBLP:conf/ih/Mittelholzer99}
and one based on complexity
theory~\cite{DBLP:conf/crypto/HopperLA02,stego1}.  The latter approach
is more concrete and has the potential of allowing more efficient
constructions. 

Most steganographic constructions supported by
provable security guarantees are instantiations of the following basic
procedure (often referred to as ``rejection-sampling''). The problem
specifies a family of message distributions (the ``channel
distributions'') that provide a number of possible options for a
so-called ``covertext'' to be transmitted. Additionally, the sender
and the receiver possess some sort of private information (typically a
keyed hash function, MAC, or other similar function) that maps channel
messages to a single bit.  In order to send a message bit $m$, the
sender draws a covertext from the channel distribution, applies the
function to the covertext and checks whether it happens to produce the
``stegotext'' $m$ she originally wished to transmit.  If this is the
case, the covertext is transmitted. In case of failure, this procedure
is repeated.
%While this is a fairly concrete procedure, there are a number of
%choices to be made with both practical and theoretical
%significance. From the security viewpoint, one is primarily interested in
%the choice of the function that is shared between the sender and the
%receiver. From a practical viewpoint, one is primarily interested in how
%the channel is implemented and whether it conforms to the various
%constraints that are imposed on it by the steganographic protocol
%specifications (e.g., are independent draws from the channel allowed?
%does the channel remember previous draws? etc.).

%As mentioned above, the security of a stegosystem can be naturally
%phrased in information-theoretic terms
%(cf.~\cite{DBLP:conf/ih/Cachin98}) or in complexity-theoretic
%terms~\cite{DBLP:conf/crypto/HopperLA02}. 

The complexity-theoretic approach to steganography considers the following experiment for the warden-adversary:
The adversary selects a message to be embedded and
receives either covertexts that embed the message or  covertexts
simply drawn from the channel distribution (without any embedding). The
adversary is then asked to distinguish between the two cases. Clearly,
if the probability of success is very close to $1/2$ it is natural to
claim that the stegosystem provides security against such
(eavesdropping) adversarial activity. Formulation of stronger attacks
(such as active attacks) is also possible.
%Observe that our model differs from the "traditional" approach to steganography where the sender modifies a covertext that is \emph{known} to the adversary in an effort to embed secret data. Such an approach is secure only against adversaries with limited detection capability. This approach is found, for instance, in several software applications which manipulate certain pixels of visual images to embed hidden information. While such minor perturbations to an image may be imperceptible to the human eye, it is trivially discerned by an algorithm with access to the original cover image. 

Given the above framework, Kiayias et
al.~\cite{DBLP:conf/ih/KiayiasRR05} (a full version appears in
\cite{stego1}) define a notion of \emph{one-time stegosystem}: this is a
steganographic protocol that is meant to be used for a single message
transmission and is proven secure in an information-theoretic sense,
provided that the key shared between the sender and the
receiver is of sufficient length. This system is a natural analogue of
a one-time pad for steganography. They then point out that this can be
used to induce a system for longer messages using standard
techniques. We shall adopt this same perspective, focusing on
achieving optimal usage of randomness.

In this paper, we present a steganography protocol that embeds a
message of length $n$ using a shared secret key of length $(1+o(1))n$
bits while achieving security $2^{-n/\polylog n}$. In this sense,
our protocol is {\bf truly randomness efficient}: the rate of key over
message approaches 1 for large values of $n$. In the previous best
known protocol \cite{DBLP:conf/ih/KiayiasRR05}, the length of the
shared secret key is at least $(2+o(1))n$ bits, regardless of the
security achieved. The key length requirement of $(2+o(1))n$ bits was
dictated by the fact that they perform ``single'' rejection sampling,
in which case some of the randomness used to interrogate the channel
during rejection sampling is discarded; as a result they must use a
$2n$-wise independent family of functions (where $n$ is the length of
the message).

Our improvement involves a number of technical elements: we introduce
the use of randomness extractors in this context and perform a variant
of rejection sampling which is more efficient in its use of the shared
secret key. However, this randomness-efficient notion of rejection
requires that we control significant new dependencies in the resulting
distribution of covertexts. Thus, while the relative improvement in
the number of random bits used by our protocol is not particularly
impressive (we save a factor $1/2$ over previous results), the
constructions seems interesting because (i.) it achieves
asymptotically optimal usage of randomness and (ii.) develops a novel
steganographic protocol. We remark, finally, that we can
permit a portion of the shared secret key to be \emph{public} while
retaining precisely $n$ private key bits. In this setting, by
separating the public and the private randomness of the shared key, we
can achieve security of $2^{-n}$. We adopt the model of channel
abstraction first defined by von Ahn~\cite{ahn04} (and also used
in~\cite{stego1}). %This model permits us to amplify the channel entropy.

At the heart of our result is the pairing  of the rejection sampling  process with  a randomness extractor. Extractors have been used widely in cryptographic applications and to the best of our knowledge, this is the first time extractors have been employed in the design of steganographic protocols.
Given our  one-time stegosystem, it is fairly straightforward now to construct provably secure steganographic encryption for longer
messages by using a pseudorandom generator (PRG) to stretch a
random seed that is shared by the sender and the receiver to
sufficient length as shown in~\cite{stego1}. The resulting stegosystem is provably secure in the
complexity theoretic model.  

\section{Preliminaries}
\label{sec:tools}

%We say that a function $\mu: \mathbb{N} \to \mathbb{R}$ is
%\emph{negligible} if for every positive polynomial $p(\cdot)$, there
%exists an $N$ such that for all $n > N$, $\mu(n) < \frac{1}{p(n)}$.
We use the notation $x \leftarrow X$ to denote sampling an element $x$ from
a distribution $X$ and the notation $x \in_R S$  to denote sampling an element $x$ uniformly
at random from a set $S$. For a function $f$ and a distribution $X$ on its domain, $f(X)$ denotes the distribution
that results from sampling $x$ from $X$ and applying $f$ to $x$. The uniform distribution on $\{0, 1\}^{d}$ is denoted by $U_{d}$. We use the notation $|s|$ to stand for the number of symbols in a string $s$. For a probability distribution $D$ with support $X$, the notation $\Pr_{D}[x]$ denotes the probability that $D$ assigns to $x \in X$. The cardinality of a set $S$ is denoted by $|S|$. The concatenation of string $s_1$ and string $s_2$ is denoted by $s_1 \circ s_2$. ``$\log$'' indicates the logarithm base 2. 

\paragraph{Pointwise $\epsilon$-biased functions} 
\begin{definition}[\cite{ahn04}]
Let $P$ be a distribution with a finite support $X$. A function $f: X
\rightarrow Y$ is said to be \emph{pointwise $\epsilon$-biased} with
respect to $P$ if $\forall y \in Y$                      
$
\left| \Pr_{x \leftarrow P}[f(x) = y] - 1/|Y| \enspace \right| < \epsilon\,.
$
\end{definition}
In this paper, we refer to such functions as $\epsilon$-biased and drop the ``pointwise'' qualification for simplicity.

\paragraph{Min-entropy} 
A distribution $X$ is said to have min-entropy of $t$ bits if the probability it assigns to  each element in its range is bounded above by $2^{-t}$.  A distribution with min-entropy at least $t$ is called a \emph{$t$-source}. %As an example of such a source, consider the distribution on $\{0, 1\}^n$ where some $n - t$ bits are fixed and the remaining $t$ bits are uniform and independent of each other. Such a distribution has min-entropy $t$. 

\begin{definition}
The \emph{min-entropy} of a random variable $X$, taking values in a
set $V$, is the quantity $H_\infty(X) \triangleq \min_{v \in V} \left(-\log \Pr[X = v]\right)$\,.
\end{definition}

\paragraph{Statistical Distance}
We use statistical distance to measure the distance between two random
variables. 
\citet{ntb} presents a detailed discussion on statistical distance and its properties.

\begin{definition}
Let $X$ and $Y$ be random variables which both take values in a finite set $S$ with probability distributions $P_{X}$ and $P_{Y}$. The statistical distance between $X$ and $Y$ is defined as 
$\Delta\left[X,Y\right] \triangleq ({1}/{2}) \sum_{s\in S}
\left|P_{X}(s)-P_{Y}(s)\right|$.
%=\max_{E \subseteq S} \left| \Pr\left[ X \in E\right] - \Pr\left[ Y \in E \right] \right|.
We say that $X$ and $Y$ are $\epsilon$-\emph{close} if $\Delta\left[X,Y\right] \leq \epsilon$. 
\end{definition}
We will use the following properties of statistical distance which follow directly from the definition.
\begin{fact}
\label{stat:prop}
Let $X$, $Y$ and $Z$ be random variables taking values in a finite set
$S$. We have (i.) $0 \leq \Delta\left[X,Y\right] \leq 1$ and (ii.) the
triangle inequality: $\Delta\left[X,Z\right] \leq \Delta\left[X,Y\right] + \Delta\left[Y,Z\right]$.
\end{fact} 

\begin{fact} [\cite{ntb}]
\label{thm:sd2}
If $S$ and $T$ are finite sets, $X$ and $Y$ are random variables
taking values in the set $S$ and $f: S \rightarrow T$ is a function, then $\Delta\left[f(X),f(Y)\right]\leq\Delta\left[X,Y\right]$.	
\end{fact}

\begin{lemma}
\label{lemma1}
Consider two random variables $(X, Y)$ and $(X', Y')$, both taking
values in $\mathcal{X} \times \mathcal{Y}$. For a particular value $x
\in \mathcal{X}$ in the support of $X$, we let $Y_x$ denote the random variable $Y$
conditioned on the event $X = x$ and define $Y'_x$ likewise. Then
$\Delta\left[\left(X,Y\right),\left(X^{\prime},Y^{\prime}\right)\right]\leq\Delta\left[X,X^{\prime}\right]+
\Exp_X \bigl[\Delta\left[Y_X,Y^{\prime}_X\right]\bigr]$.
\end{lemma}
We include the proof in Appendix~\ref{appendix_omittedproof} for
completeness.
% We also note that the statement of the lemma holds when $Y$ is dependent on $X$ and $Y^{\prime}$ is dependent on $X^{\prime}$.

\subsection{Extractors}
\label{subsec:ext}
  
Extractors are deterministic functions that operate on arbitrary
distributions with sufficient randomness and output ``almost''
uniformly distributed, independent random bits. Extractors require an
additional input: a short seed of truly random bits as a catalyst to
``extract'' randomness from such distributions. Thus the input to an
extractor contains two independent sources of randomness: the actual
distribution (the source) and the seed. Extractors were first defined
by~\citet{nz96}.
\begin{definition}
  A $(t, \epsilon)$-\emph{extractor} is a function $\Ext : \{0,
  1\}^\nu \times \{0, 1\}^d \rightarrow \{0, 1\}^\mu$ such that for
  every distribution $X$ on $\{0, 1\}^\nu$ with $H_\infty(X) \geq t$,
  the distribution $\Ext(X,U_{d})$ is $\epsilon$-close to the uniform
  distribution on $\{0, 1\}^{\mu}$.
\end{definition}

For our application, we require a stronger property from the extractor. We need the output of the extractor to remain essentially uniform even given the knowledge of the seed used. In other words, we require the extractor to extract randomness only from the source and not from the seed. A way of enforcing this condition is to demand that when the seed is concatenated to the output, the resulting distribution is still $\epsilon$-close to uniform. Such an extractor is called a \emph{strong} extractor to distinguish from the weaker notion of extractors defined above. The extractors defined above guarantee to extract randomness from $t$-sources on an average seed while strong extractors guarantee to extract randomness for most seeds. In this paper, we use the term extractor to refer to a strong extractor. 

\begin{definition} 
A $(t, \epsilon)$-strong extractor is a function $\Ext : \{0, 1\}^\nu \times \{0, 1\}^d \rightarrow \{0, 1\}^\mu$
such that for every distribution $X$ on $\{0, 1\}^\nu$ with $H_\infty(X) \geq t$, the distribution $S \circ \Ext(X,S)$ is $\epsilon$-close to the uniform distribution on $\{0, 1\}^{\mu+d}$ where $S$ is distributed according to $U_{d}$. 
\end{definition}

We refer to $\nu$ as the \emph{length of the source}, $t$ as the \emph{min-entropy threshold}, $\epsilon$ as the \emph{error} of the extractor, the ratio $t/\nu$ as the \emph{entropy rate} of the source $X$ and to the ratio $\mu/t$ as the \emph{fraction of randomness} extracted by the extractor. The entropy loss of the extractor is defined as $t + d - \mu$. The two inputs of the extractor have a total min-entropy of at least $t + d$ and the entropy loss measures how much of this randomness was ``lost'' in the extraction process. \citet{rts00} showed that no non-trivial $(t, \epsilon)$-extractor can extract all the randomness present in its inputs and must suffer an entropy loss of $2 \log(1/\epsilon) + O(1)$. For our application, we need efficient, explicit strong extractor constructions as defined below.

\begin{definition} [\cite{rssurvey}]
For functions $t(\nu)$, $\epsilon(\nu)$, $d(\nu)$, $\mu(\nu)$ a family
$\Ext = \{\Ext_\nu\}$ of functions $\Ext_\nu: \{0, 1\}^\nu \times \{0, 1\}^{d(\nu)} \rightarrow \{0, 1\}^{\mu(\nu)}$
is an explicit $(t, \epsilon)$-strong extractor if $\Ext(x, y)$ can be computed in polynomial time in its input length and for every $\nu$, $\Ext_\nu$ is a $(t(\nu), \epsilon(\nu))$-extractor.
\end{definition}

An important property of strong extractors which makes it attractive for our application is that for any $t$-source, a $(1-\epsilon)$ fraction of the seeds extract randomness from that source. 

\paragraph{Remark (\cite{rsw00}).} Let $\Ext : \{0, 1\}^\nu \times
\{0, 1\}^d \rightarrow \{0, 1\}^\mu$ be a $(t, \epsilon)$-strong
extractor. From the definition of a strong extractor, we know that $
\mathbb{E}_{s}\left[\Delta\left[\Ext(X,s),U_{\mu}\right]\right]  \leq
\epsilon $ where $s \in_R \{0,1\}^d$. By applying Markov's inequality,
we can see that \\ \mbox{$\Pr_{s}[\Delta\left[\Ext(X,s),U_{\mu}\right]
  \geq \epsilon \cdot r]\leq{1/r}$.}
%Later on in the paper, we will use this result for $r = \epsilon^{-2/3}$ and $r = \epsilon^{-1/2}$.

See the survey articles by~\citet{rssurvey, nsurvey1}, and~\citet{nsurvey2}  for more details on extractors and their properties. In this paper, we use the explicit strong extractor construction by Raz, Reingold and Vadhan \cite{rrv99} which works on sources of any min-entropy. It extracts all the min-entropy using $O(\log^3 \nu)$ additional random seed bits while achieving an optimal entropy loss (up to an additive constant) of $\chi = 2 \log (1/\epsilon) + O(1)$ bits. 

\begin{theorem}[RRV Extractor \cite{rrv99}]
\label{rrvcons}
For every $\nu$, $t \in \mathbb{N}$, and $\epsilon > 0$  such that $t \leq \nu$, there are explicit $(t, \epsilon)$-strong extractors 
$\Ext: \{0,1\}^\nu \times \{0,1\}^d \rightarrow \{0,1\}^{t-\chi}$ 
with entropy loss $\chi = 2 \log (1/\epsilon) + O(1)$ bits and
requiring seeds of length
\[
d = O(\log^2 \nu \cdot \log(1/\epsilon) \cdot \log t) \textrm{ bits}.
\]
\end{theorem}

\subsection{The channel model}
The security of a steganography protocol is measured by the adversary's ability to distinguish between ``normal'' and ``covert'' message distributions over a communication channel. To characterize normal communication we define and formalize the \emph{communication channel} following standard terminology used in the literature \cite{DBLP:conf/crypto/HopperLA02, DBLP:conf/ih/Cachin98, ahn04, stego1, DBLP:journals/tc/HopperAL09}. We let $\Sigma$ denote the symbols of an alphabet and treat the \emph{channel} as a family of distributions $\mathcal{C} = \{{C_h}\}_{h \in \Sigma^\ast}$; each ${C_h}$ is supported on $\Sigma$.  These channel distributions model a history-dependent notion of channel data. 
%We let $C_h$ denote the marginal channel distribution on a single symbol from $\Sigma$ conditioned on the history $h$ of previously drawn symbols and $C_{h}^t$ denotes the marginal distribution on sequences of $t$ symbols conditioned on history $h$. 

We adopt the model of channel abstraction first defined
by~\citet{ahn04}. Here, Alice is provided with a means for sampling
``deep into the channel.'' In particular, Alice and, consequently, the
steganographic encoding protocol, has access to a channel oracle that
can sample from the channel for \emph{any} history. Formally, during
the embedding process, Alice may sample from $C_{h_1\circ \ldots \circ
  h_\ell}$ for any history she wishes (though Alice is constrained to
be efficient and so can make no more than polynomially many queries of
polynomial length). This model allows Alice to transform a channel $C$
with min-entropy $\delta$ into a channel $C^{\pi}$ with min-entropy
$\pi \delta$. Specifically, the channel $C^{\pi}$ is defined over the
alphabet $\Sigma^\pi$, whose elements we write as vectors $\mathbf{h}
= (h_1, \ldots, h_\pi)$. The distribution $C^{\pi}_{\mathbf{h}^1,
  \ldots, \mathbf{h}^v}$ is determined by the channel $C$ with history
$h^1_1 \circ \cdots \circ h_\pi^1 \circ h^2_1 \circ \cdots \circ
h^v_\pi$. This definition captures the adaptive nature of the channel
by taking into account the dependence between symbols as is typical in
real world communications. We assume that the channel satisfies a
min-entropy constraint for all histories. We say that a channel has
min-entropy $\delta$ if $\forall h \in \Sigma^\ast$, $H_\infty(C_h)
\geq \delta$. Observe that this implies that $H_\infty(C_h^{\pi}) \geq
\delta \pi$ due to the additive nature of marginal min-entropy.
  
\subsection{One-time stegosystem}
\label{sec:stegodef}
Here, we give the definition of a \emph{one-time stegosystem}, a steganographic
system that enables the one-time steganographic transmission of a message
provided that the two parties share a suitable key. We adopt the definitions used by~\citet{stego1}.

\begin{definition}
A \emph{one-time stegosystem} consists of three probabilistic polynomial time 
algorithms $S=(SK, SE, SD)$, where:
\begin{itemize}
\item $SK$ is the \emph{key generation algorithm}; we write $SK(1^k)=\kappa$. It produces a key $\kappa$ of length $k$. 
\item $SE$ is the \emph{embedding procedure} and has access to the channel; $SE\left(\kappa,m;\mathcal{O}\right) = s \in \Sigma^*$. The embedding procedure takes into account the history $h$ of communication that has taken place between Alice and Bob thus far and begins its operation corresponding to this history. It takes as input the key $\kappa$ of length $k$, a message $m$ of length $n = n(k)$ and accesses the channel through 	an (probabilistic) oracle $\mathcal{O}$. The oracle $\mathcal{O}$ accepts as input \emph{any} polynomial length history $h^{\prime} \in \Sigma^{\ast}$ and allows $SE$ to draw independent samples repeatedly from $C_{h \circ h^{\prime}}$. The output is the stegotext $s \in \Sigma^*$. Observe that in a one-time stegosystem, once a security parameter $k$ is chosen, the length of the message $n$ is a fixed function of $k$. In our model of channel abstraction, $SE$ can access the channel for \emph{any} history.
\item $SD$ is the \emph{extraction procedure}; $SD\left(\kappa,c \right)= m$. It takes as input the key $\kappa$ of length $k$, and some $c \in \Sigma^\ast$. The output is a message $m$.
\end{itemize}
\end{definition}

\noindent We next define a notion of correctness for a one-time stegosystem.

\begin{definition}[Correctness] A one-time stegosystem $(SK, SE, SD)$ is said to be \emph{$(\epsilon,\delta)$-correct} provided that for all channels $\mathcal{C}$ of min-entropy $\delta$, 
it holds that $\forall h \in \Sigma^{\ast}$
\[ 
 \forall m \in\{0,1\}^{n(k)} \,\, \Pr[SD(\kappa, SE(\kappa, m;\mathcal{O})) \neq m \mid \kappa \gets SK(1^k)  ]
\leq \epsilon ~.
\]
\end{definition}

In general, we treat both $\epsilon = \epsilon(k)$ and $\delta = \delta(k)$ as functions of $k$, the security parameter and the oracle $\mathcal{O}$ as a function of the history $h$. 

One-time stegosystem security is based on the indistinguishability between a transmission
that contains a steganographically embedded message and a transmission that
contains no embedded messages. The adversarial game discussed next is meant to model the behavior of a warden in the Simmons' formulation of the problem discussed earlier.

An adversary $\mathcal{A}$ against a one-time stegosystem $S = (SK,
SE, SD)$ is a pair of algorithms $\mathcal{A}=(SA_1, SA_2)$, that
plays the following game, denoted $G^\mathcal{A}(1^k)$:
\begin{enumerate}
\item  A key $\kappa$ is generated by $SK(1^k)$.
  
\item Algorithm $SA_1$ receives as input the security parameter $k$ and outputs a triple $(m^\ast, \textrm{aux}, h_{\sf c}) \in M_{n} \times \{0,1\}^\ast \times \Sigma^*$, where $\textrm{aux}$ is some auxiliary information that will be passed to $SA_2$. $SA_1$ is provided access to the channel via the oracle $\mathcal{O}$, which takes the history $h$ as input. The channel oracle $\mathcal{O}$, on input $h$, returns to $SA_1$ an element $c$ selected according to $C_h$. This way, the adversary can learn about the channel distribution for \emph{any} history.
\item A bit $b$ is chosen uniformly at random. 
\begin{itemize}
\item If $b = 0$, let $c^\ast \gets SE(\kappa,m^\ast;\mathcal{O})$, so $c^\ast$ is a stegotext. 
\item If $b = 1$, let $c^\ast = c_1 \circ \cdots \circ c_\lambda$ where $\lambda = | SE(\kappa,m^{\ast};\mathcal{O})|$ and \mbox{$c_i \stackrel{r}{\gets} C_{h \circ \sf c_1 \circ \cdots \circ c_{i-1}}$.} In this case, $c^\ast \gets C_h^{\lambda}$.

\end{itemize}
\item The input for $SA_2$ is $1^k$, $h_{\sf c}$, $c^\ast$ and $\textrm{aux}$. $SA_2$ outputs a bit $b'$. If $b' = b$ then we say that $(SA_1, SA_2)$ \emph{succeeded} and write $G^{\mc{A}}(1^k) = \text{success}$.
\end{enumerate}
The \emph{advantage} of the adversary $\mc{A}$ over a stegosystem $S$
is defined as: $\mbox{{\bf Adv}}_S^\mathcal{A}(k) =  \left|\Pr\big[ G^\mathcal{A}(1^k) = \text{success}  \big ] - {1}/{2} \right|$.
 
 The probability includes the coin tosses of $\mathcal{A}$ and $SE$.
 %as well as the coin tosses of $G^\mathcal{A}(1^k)$. 
 The (information-theoretic) insecurity of the stegosystem is defined as
\[
\mbox{\bf{InSec}}_{S}(k) =  \max_{\mathcal{A} }\{\mbox{\bf{Adv}}_S^\mathcal{A}(k)\}\,,
\]
this maximum taken over all (time unbounded) adversaries $\mc{A}$.
\begin{definition}[Security]
We say that a stegosystem is \emph{$(\epsilon, \delta)$-secure} if for all channels with min-entropy $\delta$ we have $\mbox{\bf{InSec}}_{S}(k) \leq \epsilon$.
\end{definition}

\paragraph{Overhead.}
The {\em overhead} of a one-time stegosystem is judged by the relation of the key length $k$ and message length $n$. We adopt the ratio $k/n$ as the measure of overhead as first defined by~\citet{stego1}. 

\subsection{Rejection Sampling}
\label{subsec:rejsam}

As noted before, a common method used in steganography employing a channel distribution
is that of \emph{rejection sampling} (cf. \cite{DBLP:conf/ih/Cachin98,DBLP:conf/crypto/HopperLA02,stego1}).
We use a variant of rejection sampling to transmit bit vectors as
opposed to a single bit. To transmit bit vectors, we amplify the entropy of the channel as discussed before and apply $\rho$-rejection sampling described below. More precisely, we transform a channel $C$ with
min-entropy $\delta$ into a channel $C^{\pi}$ with min-entropy $\pi
\delta$, defined over the alphabet $\Sigma^\pi$. We now perform $\rho$-rejection sampling over $C^{\pi}$ as described: Assuming that one wishes to transmit a bit vector $\vec{m} \in \{0,1\}^\eta$ and employs a random function $f:\Sigma^{\pi}\to\{0,1\}^{\eta}$, one performs the following {}``rejection sampling'' process: 
\begin{center}
\begin{tabular}{|l|}
\hline 
$\rejsam_{h}^{f}(\vec{m},\rho)$\\
\hline 
\texttt{let $j=0$}\\
\hspace{1 cm}\texttt{repeat:}\\
\hspace{2 cm}\texttt{sample }$\vec{c}\gets$ $C_{h}^{\pi}$
, \texttt{increment $j$}\\					
\hspace{1 cm}\texttt{until }$f(\vec{c})=\vec{m}$ or $(j>\rho)$\\
\texttt{output: $\vec{c}$} \\
\hline
\end{tabular}
\end{center}
For a given history $h$, the procedure $\rejsam_{h}^{f}(\vec{m},\rho)$ draws independent samples from the channel distribution $C_h^\pi$ in rounds until $f(\vec{c})=\vec{m}$ or $j>\rho$. As there are at most a total of $\rho+1$ rounds, if none of the first $\rho$ samples drawn map to the target bit vector, the sample drawn at round $\rho+1$ is returned by the procedure. Here, as defined before, $\Sigma^\pi$ denotes the output alphabet of the channel, $h$ denotes the
history of the channel at the start of the process, and
$C^{\pi}_{h}$ denotes the marginal distribution on sequences of $\pi$ symbols given by the channel after 
history $h$.  The receiver (also privy to the function $f$)
applies the function to the received message $\vec{c} \in \Sigma^{\pi}$ and recovers $\vec{m}$
with a certain probability of success.  
Note that the above process performs $\rho+1$ draws from the channel
with the \emph{same} history. These draws are assumed to be independent. One basic property of rejection sampling that we use is:

\begin{lemma}[\cite{ahn04}]
\label{lem:rej-samp}
If the function $f$ is $\epsilon$-biased on $C^\pi_h$ for history $h$, then for any $\rho$ and uniformly random $\vec{m}\in_R\{0,1\}^{\eta}$:
$$
\Delta\left[\rejsam_{h}^{f}(\vec{m},\rho),C^\pi_h\right] \leq \epsilon.
$$
\end{lemma}
\begin{proof}
Let us denote the samples drawn by the procedure $\rejsam_{h}^{f}(\vec{m},\rho)$ as $\vec{c_i}, i=1,\cdots,\rho+1$. Suppose the target bit vector $\vec{m}$ was chosen with the probability $P_f^{(\vec{m})} \triangleq \Pr[f(C_h^\pi)=\vec{m}]$, i.e, $\vec{m} \gets P_f^{(\vec{m})}$, we first show that the output from $\rejsam_{h}^{f}(\vec{m},\rho)$ is distributed identically to $C_h^\pi$. For simplicity of notation, let us define $p_{m} \triangleq \Pr_{P_f^{(\vec{m})}}[\vec{m}]$. Let $p_c$ denote the probability of drawing $\vec{c}$ from the channel distribution $C_h^\pi$, i.e., $p_c \triangleq \Pr_{C_h^\pi}[\vec{c}]$. For $\vec{c} \in C_h^\pi$, the probability of observing $\vec{c}$ under the $\rejsam_{h}^{f}(\vec{m},\rho)$ procedure is then given by \\~\\
$\Pr[\rejsam_{h}^{f}(\vec{m},\rho)=\vec{c}]$
\begin{eqnarray*}
& = & \Pr_{_{\vec{c}_{1}\gets C_{h}^{\pi}}}[\vec{c}_{1}=\vec{c}]\cdot\Pr[f(\vec{c}_{1})=\vec{m}]+\textrm{\ensuremath{\Pr_{_{\vec{c}_{2}\gets C_{h}^{\pi}}}}}[\vec{c}_{2}=\vec{c}]\cdot\Pr[f(\vec{c}_{2})=\vec{m}]\cdot\Pr[f(\vec{c}_{1})\neq\vec{m}]\\
 &  & +\Pr_{_{\vec{c}_{3}\gets C_{h}^{t}}}[\vec{c}_{3}=\vec{c}]\cdot\Pr[f(\vec{c}_{3})=\vec{m}]\cdot\Pr[f(\vec{c}_{1})\neq\vec{m}\wedge f(\vec{c}_{2})\neq\vec{m}]+\cdots\\
 & = & p_{c}p_{m}+p_{c}p_{m}\left(1-p_{m}\right)+\cdots+p_{c}p_{m}\left(1-p_{m}\right)^{\rho-1}+p_{c}\left(1-p_{m}\right)^{\rho}\\
 &  & \\
% & = & p_{c}p_{m}\left(1+\left(1-p_{m}\right)+\left(1-p_{m}\right)^{2}+\cdots+\left(1-p_{m}\right)^{\rho-1}\right)+p_{c}\left(1-p_{m}\right)^{\rho}\\
 & = & p_{c}p_{m}\left(\frac{1-\left(1-p_{m}\right)^{\rho}}{p_{m}}\right)+p_{c}\left(1-p_{m}\right)^{\rho} = p_{c}.
\end{eqnarray*}
From the above discussion, we can see that when the target bit vector $\vec{m}$ was chosen from the distribution $P_f^{(\vec{m})}$, the output from $\rejsam_{h}^{f}(\vec{m},\rho)$ is distributed identically to $C_h^\pi$. Since $f$ is $\epsilon$-biased, $\Delta\left[U_{\eta}, P_f^{(\vec{m})}\right] \leq \epsilon$.  Hence, $$\Delta\left[ \rejsam_{h}^{f}(\vec{m} \gets U_{\eta},\rho), \rejsam_{h}^{f}(\vec{m} \gets P_f^{(\vec{m})},\rho)\right] \leq \epsilon$$ by Fact~\ref{thm:sd2} which gives us the statement of the lemma.
\end{proof}

\section{The construction}

In this section, we outline our construction of a one-time stegosystem
as an interaction between Alice (the sender) and Bob (the receiver).
Alice and Bob wish to communicate over a channel $C_h^{\pi}$ with history $h$. We also assume that the support of $\mathcal{C}_h$ is $\{0,1\}^b$, i.e, $|\Sigma|=2^b$. 

\subsection{A one-time stegosystem} 
\label{sec:stegosystem}

Let $\vec{m} \in \{0,1\}^{n}$ be the message to be embedded. Our stegosystem uses the \emph{RRV} strong-extractor construction as described in Theorem \ref{rrvcons} which extracts randomness from the distribution $C_h^\pi$ supported on $\{0,1\}^{\pi \cdot b}$ by rejection sampling as described in Section \ref{subsec:rejsam}. Specifically, we will use the extractor with the seed $s$ as the function $f$ in the rejection sampling procedure.

\noindent Alice and Bob agree on the following:
\begin{sloppypar}
\begin{description}
\item[Extractor Construction.] Alice and Bob agree to use the explicit \emph{RRV} strong-extractor construction as described in Theorem \ref{rrvcons}. They use a seed $s \in_{R} \{0,1\}^{d}$ for the extractor. The length of the seed $d$ will be determined later as a function of $\delta, n, b$ and security $\epsilon$. The notation $E_{s}$ stands for the extractor used with the seed $s$ i.e., $E(\cdot,s)$. Here, we treat the seed $s$ as private and in Section \ref{sec:randopt} we show that the seed $s$ may be  \emph{public} and discuss the implications of this choice. 

\item[One-Time Pad.] Alice and Bob also use a shared one-time pad secret key $\kappa^{\text{otp}} \in_R \{0,1\}^{n}$ effectively transmitting $\vec{m}^{\prime}=\kappa^{\text{otp}}\oplus\vec{m}$.
\item[Shared Secret Key.] The secret key that they now share is $\kappa = (\kappa^{\text{otp}},s)$ of length $k = n+d$.
\end{description}
\end{sloppypar}
Key generation consists of generating the one-time pad secret key $\kappa^{\text{otp}} \in_R \{0,1\}^{n}$ and the random seed $s$ of length $d$ to be used with the extractor. The encoding procedure accepts an input message $\vec{m}$ of length $n$ bits and outputs a stegotext of length $\lambda$. We will analyze the stegosystem below in terms of the parameters $\pi$, $d$, $\lambda$, $\rho$ and some constant $c>1$ relegating discussion of how these parameters determine the overall efficiency of the system to Section~\ref{sec:parameters}.
\begin{figure}[h]
\begin{centering}
\begin{tabular}{@{}|l|l|}
\hline 
\label{protocol} 
\texttt{PROCEDURE $SE$: } & \texttt{PROCEDURE} $SD$: \tabularnewline 
\hline 
%{} & {}
%\tabularnewline 
\texttt{Input: Key $\kappa = (\kappa^{\text{otp}},s)$; $\vec{m}\in\left\{ 0,1\right\} ^{n}$,} & \texttt{Input: Key $\kappa = (\kappa^{\text{otp}},s)$}
\tabularnewline
\hspace{1.45cm}\texttt{history $h\in\Sigma^{*}$ } & \hspace{1.45cm}\texttt{stegotext $\sf{c_{stego}}$ }
\tabularnewline
\texttt{let $\vec{m}^{\prime}=\kappa^{\text{otp}}\oplus\vec{m}$ } & {}
\tabularnewline
\texttt{parse $\vec{m}^{\prime}$ as $\vec{m}^{\prime}=\vec{m^{\prime}_1}\vec{m^{\prime}_2}\dots\vec{m^{\prime}_{\lceil \textrm{$n$}/\textrm{$c$}\log \textrm{$n$} \rceil}}$ } & \texttt{parse: $\sf{c_{stego}}=\vec{c_1}\vec{c_2}\ldots\vec{c_{\lceil \textrm{$n$} /\textrm{$c$}\log \textrm{$n$} \rceil}}$ }
\tabularnewline
\texttt{for $i=1$ to }$\lceil n/c\log n \rceil$\texttt{ \{ } & \texttt{for $i=1$ to $\lceil n /c\log n \rceil$ do \{}
\tabularnewline
\hspace{1 cm}$\vec{c_i}\leftarrow \rejsam_{h}^{E_{s}}(\vec{m_i^{\prime}},\rho)$ & \hspace{1 cm} \texttt{set $\vec{m_{i}}^{\prime}=E_{s}(\vec{c}_{i})$ }
\tabularnewline
\hspace{1 cm} \texttt{set $h\gets h\circ \vec{c}_{i}$ } & \texttt{\}}
\tabularnewline
\}  & \texttt{set $\vec{m}^{\prime}=\vec{m^{\prime}_1}\vec{m^{\prime}_2}\ldots\vec{m^{\prime}_{\lceil \textrm{$n$} /\textrm{$c$}\log \textrm{$n$} \rceil}}$}
\tabularnewline
\texttt{Output: $\sf{c_{stego}}=\vec{c_1}\vec{c_2}\ldots\vec{c_{\lceil \textrm{$n$} /\textrm{$c$}\log \textrm{$n$} \rceil}}\in\Sigma^{\lambda}$} & \texttt{Output: $\vec{m}^{\prime}\oplus \kappa^{\text{otp}}$ }
%\tabularnewline
%{} & {}
\tabularnewline 
\hline 
\end{tabular}
%\par			
\end{centering}
\caption{Encryption and Decryption algorithms for the one-time stegosystem of \ref{protocol}.}
\label{fig:protocol} 
\end{figure}

Alice and Bob communicate using the algorithm $SE$ for steganographic embedding
and $SD$ for decoding as described in Figure~\ref{fig:protocol}. In $SE$, after applying the one-time pad to randomize her message $\vec{m}$, Alice obtains $\vec{m^{\prime}}=\kappa^{\text{otp}}\oplus\vec{m}$. She then parses $\vec{m^{\prime}}$ into $\lceil n/c\log n \rceil$ blocks, each block of length $c \log n$ for some constant $c > 1$, i.e., $\vec{m^{\prime}} = \vec{m_{1}}^{\prime}\vec{m_{2}}^{\prime}\dots\vec{m}_{\lceil n/c\log n \rceil}^{\prime}$. She then applies the procedure $\rejsam^{E_{s}}_{h}(\vec{m}_i^{\prime},\rho)$ to obtain an element $\vec{c_i} \in \Sigma^\pi$ for each block $\vec{m_i}^{\prime}, i = 1, \cdots, \lceil n/c\log n \rceil$ of the randomized message. Here, the history $h$ represents the current history at the time of the rejection sampling procedure which is updated after the completion of the procedure. Recall that the notation $E_{s}$ stands for the extractor used with the seed $s$ i.e., $E(\cdot,s)$. The resulting stegotext, denoted by $c_{\text{stego}}$ that is transmitted to Bob is $c_{\text{stego}} = \vec{c_{1}}\vec{c_{2}}\ldots\vec{c}_{\lceil n/c\log n\rceil}$. In $SD$, the received stegotext is first parsed into $\lceil n/c\log n \rceil$ blocks as shown and then evaluated using the extractor with seed $s$ for each block; this results in a message block. After performing this for each received block, a bit string of length $n$ is obtained, which is subjected to the one-time pad decoding to obtain the original message. The detailed security and correctness analysis follow in the next two sections.

\subsection{Security}
\label{sec:security}

In this section, we argue about the security of our one-time stegosystem. Specifically, we establish an upper bound on the statistical distance between the ``normal'' and ``covert'' message distributions over the communication channel. First, by Lemma \ref{lem:rej-samp}, observe that if the function $f$ is $\epsilon$-biased on $C_{h}^{\pi}$ for history $h$, then for any $\rho$, $\vec{m}^{\prime}\in_R\{0,1\}^{\eta}$: $\Delta[\rejsam_{h}^{f}(\vec{m}^{\prime},\rho),C_{h}^{\pi}] \leq \epsilon$. Now, consider the strong extractor $\Ext : \{0, 1\}^\nu \times \{0, 1\}^d \rightarrow \{0, 1\}^\mu$ used in the rejection sampling procedure. Denote the error of extractor by $\epsilon_{ext}$. Recall from the remark in Section \ref{subsec:ext} that, for a uniformly chosen seed $s \in_R \{0,1\}^{d}$, $\Pr_{s}[\Delta\left[\Ext(X,s), U_{\mu}\right] \geq {\sqrt{\epsilon_{ext}}}] \leq {\sqrt{\epsilon_{ext}}}$. From this we can see that $\Ext$ fails to be a $\sqrt{\epsilon_{ext}}$-biased function with probability no more than ${\sqrt{\epsilon_{ext}}}$ in the choice of the seed $s$. Thus, for a random $m^{\prime}$ and $s$, 
\[
\Delta[\rejsam_{h}^{E_{s}}(\vec{m}^{\prime},\rho),C_{h}^{\pi}] \leq 1\cdot\sqrt{\epsilon_{ext}}+{\sqrt{\epsilon_{ext}}}\cdot 1 \leq 2\sqrt{\epsilon_{ext}}\,.
\]
We obtain the above inequality by upper bounding the probability of
the extractor being a $\sqrt{\epsilon_{ext}}$-biased function by $1$
and observing that the statistical distance is also upper bounded by 1
by Fact~\ref{stat:prop}. Suppose that in our stegosystem construction,
we had used an independent and uniformly chosen seed $s_i \in_R
\{0,1\}^{d}$ for each message block $i = 1, 2, \cdots, \lceil n/c\log
n \rceil$, the statistical distance between $C_{h}^{\lambda}$ and the
output of the procedure $SE$ would then be 
\[
\Delta\left[SE(\kappa, \vec{m};\mathcal{O}),C_{h}^{\lambda}\right] \leq 2\sqrt{\epsilon_{ext}} \lceil n/ c \log n \rceil.
\]
However, employing an independent and uniformly chosen seed for each
message block would require too much randomness. In our scheme, we
employ \emph{a single seed} $s$ over all the message blocks and so we
need to manage the dependencies between the output covertexts; this is
the major technical issue in the proof, which is relegated to
Appendix~\ref{appendix_secproof} for lack of space. In particular, for any message
$\vec{m} \in \{0,1\}^n$, we present an upper bound on
$\Delta\left[SE(\kappa, \vec{m};\mathcal{O}),C_{h}^{\lambda}\right]$
when using a single seed $s \in_R \{0,1\}^{d}$ over all the message
blocks. We record the theorem below; the proof appears in
Appendix~\ref{appendix_secproof}.
 
\begin{theorem}
\label{thm:security}
For any $\epsilon,\delta>0$, message $\vec{m}\in\left\{ 0,1\right\}^{n}$ consider the stegosystem $(SK,SE,SD)$ of Section~\ref{sec:stegosystem} under the  parameter constraint  $\epsilon_{ext} \leq \left(\frac{\epsilon}{3 \ell}\right)^3$.
Then it holds that the stegosystem is $(\epsilon,\delta)$-secure where $\epsilon_{ext}$ is the extractor error and $\ell=\lceil n/c\log n\rceil$ for some constant $c>1$.
\end{theorem}

\subsection{Correctness}
\label{sec:soundness}

In this section we obtain an upper bound on the soundness of our stegosystem. We focus on the mapping between $\{0,1\}^n$ and $\Sigma^\lambda$ determined by the $SE$ procedure of the one-time stegosystem.  We would like to bound the probability of the stego decoding procedure's inability to faithfully recover the encoded message. 

\begin{theorem}
\label{thm:correctness}
For any $\epsilon,\delta > 0$, message $\vec{m}\in\left\{ 0,1\right\}^{n}$ consider the stegosystem $(SK,SE,SD)$ of Section~\ref{sec:stegosystem} under the parameter constraints $\epsilon_{ext} \leq \left(\frac{\epsilon}{6 \ell^2}\right)^3$ and $\rho \geq 2n^{c}\log(3 \ell \epsilon^{-1})$ for some constant $c>1$. Then it holds that the stegosystem is $(\epsilon,\delta)$-correct where $\epsilon_{ext}$ is the extractor error and $\ell=\lceil n/c \log n \rceil$ for some constant $c>1$.
\end{theorem}

\begin{proof}
Recall that the first step of the procedure $SE$ is to randomize the message $\vec{m}$ to get $\vec{m^{\prime}}=\vec{m}\oplus \kappa^{\text{otp}}$. $SE$ then proceeds to parse $\vec{m^{\prime}}$ into blocks: $\vec{m^{\prime}} = \vec{m_1^{\prime}}\vec{m_2^{\prime}}\dots\vec{m_\ell^{\prime}}$, $\ell = {\lceil n/c\log n \rceil}$. Let $F$ be the event that $SD$ is unable to correctly decode the message encoded by $SE$. We seek to upper bound the probability of $F$. We proceed to first estimate the probability of failure for one message block $\vec{m_i}$. Let us denote this event by $F^{\prime}$.

Recall that we pick a seed $s\in_{R}\left\{ 0,1\right\}^{d}$ for the
extractor we use in our construction and let $\epsilon_{ext}$ denote
the error of the extractor. We say that a seed $s$ is \emph{good} if
$\forall\tau,\enskip\mu\left(G_{s}^{\tau}\right)\geq1-\sqrt[3]{\epsilon_{ext}}$,
$\tau=1,2,\cdots,\ell$.  We show in Appendix~\ref{appendix_secproof}
that the probability of seed $s$ to be good is given by
$\Pr_{s}\left[\forall
  \tau\mid\mu\left(G_{s}^{\tau}\right)\geq1-\sqrt[3]{\epsilon_{ext}}\right]\geq1-\ell\sqrt[3]{\epsilon_{ext}}$. (This
follows from straightforward applications of Markov's inequality.)
Thus that the probability that the seed $s$ is not good is no more
than $\ell\sqrt[3]{\epsilon_{ext}}$.  By the union bound this yields
\begin{eqnarray*}
  \Pr[F] & = & \ell\cdot\left(\Pr[F^{\prime}\mid \textrm{$s$
      good}]\cdot\Pr[\textrm{$s$ good}]+\Pr[F^{\prime}\mid \textrm{$s$ not good}]\cdot\Pr[\textrm{$s$ not good}]\right)\\
  & \leq & \ell\cdot\left(\Pr[F^{\prime}\mid \textrm{$a$ good}]\cdot1+1\cdot\left(\ell\sqrt[3]{\epsilon_{ext}}\right)\right).
\end{eqnarray*}
We proceed to bound $\Pr[F^{\prime}\left|\right.s\textrm{ is good}]$. We know that when the seed $s$ is good, for no more than $\sqrt[3]{\epsilon_{ext}}$ fraction of distributions in every level $\tau=1,2,\cdots,\ell$, the extractor coupled with the seed $s$ is not a $\sqrt[3]{\epsilon_{ext}}$-biased function with probability no more than $\sqrt[3]{\epsilon_{ext}^2}$. So, we get
\[
\Pr[F^{\prime}\mid \textrm{$s$ good}]\leq 1 \cdot\left(1-\left(\frac{1}{2^{\left|\vec{m}_i\right|}}-\sqrt[3]{\epsilon_{ext}}\right)\right)^{\rho}+\sqrt[3]{\epsilon_{ext}}\cdot 1 + \sqrt[3]{\epsilon_{ext}^2} \cdot 1
\]
where $\rho$ is the bound on the number of iterations performed by the
rejection sampling procedure. Setting $\epsilon_{ext} \leq
{1}/{(8\cdot2^{3\left|\vec{m_i}\right|})}= {1}/{(8 \cdot n^{3c})}$
and
$$\rho = 2 \cdot 2^{\left|\vec{m_i}\right|}\cdot\log(3 \ell
\epsilon^{-1})=2 n^{c}\log(3 \ell \epsilon^{-1})
$$
(since in our construction $\left|\vec{m_i}\right|=c\log n$, and as $\rho$ is exponential in the block length, we choose the message block length to be $c \log n$), we have
\[
\Pr[F^{\prime}\mid \textrm{$s$ good}]\leq \frac{\epsilon}{3 \ell}+2 \sqrt[3]{\epsilon_{ext}}\,.
\]
From the statement of the theorem we have that $\epsilon_{ext} \leq
\left(\frac{\epsilon}{6 \ell^2}\right)^3$ and hence
\[
\Pr[F]  \leq  \ell\cdot\left(\Pr[F^{\prime}\left|\right. \textrm{$s$ good}]\cdot1+1\cdot\left(\ell\sqrt[3]{\epsilon_{ext}}\right)\right)
 \leq  \epsilon
\]
and the statement of the theorem follows. 
\end{proof}

%We record the following theorem for completeness.
\noindent We record the security and correctness theorem below.
\begin{theorem}
\label{thm:complete}
For any $\epsilon_{ext} \leq 1/8n^{3c}, \delta>0$, message $\vec{m}\in\left\{ 0,1\right\}^{n}$, and $\rho \geq 2n^{c}\log(\epsilon_{ext}^{-1/3})$, the stegosystem $(SK,SE,SD)$ of Section~\ref{sec:stegosystem} is $(\epsilon_{cor},\delta)$-correct and $(\epsilon_{sec},\delta)$-secure, where $\epsilon_{cor} \leq 4\ell^2 \sqrt[3]{\epsilon_{ext}}$ and $\epsilon_{sec} \leq 3\ell\sqrt[3]{\epsilon_{ext}}$. Here, $\epsilon_{ext}$ is the extractor error and $\ell=\lceil n/c\log n\rceil$ for some constant $c>1$. 
\end{theorem}

\subsection{Putting it all together}
\label{sec:parameters}

The objective of this section is to integrate the results of the
previous sections of the paper. We first show that our steganography
protocol embeds a message of length $n$ bits using a shared secret key
of length $(1+o(1))n$ bits while achieving security $2^{-n/\polylog n}$. In this sense, our protocol is randomness efficient in the shared key. We next show that by permitting a portion of the shared secret key to be \emph{public} while retaining $n$ private key bits, we can  achieve security of $2^{-n}$. Let us first start our discussion by considering the parameters of the extractor construction we employ in our protocol.

\subsubsection{Extractor Parameters} 

Recall that $\pi$ is the parameter that dictates how many copies of the channel Alice decides to use in order to transform the channel $C$ with 
min-entropy $\delta$ into a channel $C^{\pi}$ with min-entropy $\pi
\delta$. If we let \mbox{$\pi=\delta^{-1}\cdot\left(c\log n+2\log\left(1/\epsilon_{ext}\right)+O\left(1\right)\right)$} for some constant $c>1$, the channel distribution $C_h^\pi$ supported on $\{0,1\}^{\delta^{-1}\cdot\left(c\log  n+2\log\left(1/\epsilon_{ext}\right)+O\left(1\right)\right)\cdot b}$ has a min-entropy of at least $t = c\log n+2\log\left(1/\epsilon_{ext}\right)+O\left(1\right)$. To put this all together, the \emph{RRV} strong-extractor is a function
$\Ext: \left\{0,1\right\}^{\nu} \times \left\{0,1\right\}^{d} \rightarrow \left\{0,1\right\}^{t-\Delta}$
where 
\begin{eqnarray*}
\nu & = & \delta^{-1}\cdot\left(c\log n+2\log\left(1/\epsilon_{ext}\right)+O\left(1\right)\right)\cdot b \\
d & = & O\left(\log^{2}\left(\delta^{-1}\cdot\left(c\log n+2\log\left(1/\epsilon_{ext}\right)+O\left(1\right)\right)\cdot b\right)\cdot\log\left(1/\epsilon_{ext}\right)\cdot\log t\right) \\
t & = & c\log n+2\log\left(1/\epsilon_{ext}\right)+O\left(1\right) \\
\Delta & = & 2\log\left(1/\epsilon_{ext}\right)+O\left(1\right) \textrm{ and} \\ 
t - \Delta & = & c\log n
\end{eqnarray*}

We can immediately see from the preceding discussion that our stegotext is of length  $$\frac{n}{c\log n}\cdot\delta^{-1}\cdot\left(c\log n+2\log\left(1/\epsilon_{ext}\right)+O\left(1\right)\right)\cdot b = \frac{n}{\delta}\left(1+\frac{2\log\left(1/\epsilon_{ext}\right)}{c\log n}+o\left(1\right)\right)\cdot b$$
bits to embed $n$ bits of message. 

\subsubsection{Randomness Efficiency}
\label{sec:randopt}

Recall that the shared secret key between Alice and Bob is comprised
of the one-time pad $\kappa^{\text{otp}} \in_R \{0,1\}^{n}$ of length
$n$ and the extractor seed $s \in_R \{0,1\}^{d}$ of length $d$ bits,
i.e., $\kappa = (\kappa^{\text{otp}},s)$. Also, the length of the seed
from the above discussion is given by 
\[
d =  O\left(\log^{2}\left(\delta^{-1}\cdot\left(c\log
      n+2\log\left(1/\epsilon_{ext}\right)+O\left(1\right)\right)\cdot
    b\right)\cdot\log\left(1/\epsilon_{ext}\right)\cdot\log
  t\right)\,.
\]
Notice the relationship between the error of the extractor $\epsilon_{ext}$ and the desired security from our stegosystem $\epsilon$ is given by $\epsilon_{ext} \leq \left(\frac{\epsilon}{3 \ell}\right)^3$ from Theorem \ref{thm:security}. When we let $\epsilon = 2^{-n/\log^{O(1)}n}$, we can see that the length of the seed $d=o(n)$. Thus we can embed a message of length $n$ bits using a shared secret key of length $(1+o(1))n$ bits while achieving security $2^{-n/\log^{O(1)}n}$. Suppose, we were to let the extractor seed of length $d$ be public, observe now that we can attain $\epsilon = 2^{-n}$ security in the length of the shared private key of length $n$. The seed length can now be given by $d = O(n \log n \log^2(\delta^{-1}bn))$. For small $\epsilon$, the relationship between the seed length $d$ and security $\epsilon$ can be given by $d=O\left(\log^3\left(\log\left(\epsilon^{-3}\right)\right)\log\left(\epsilon^{-3}\right)\right)$. We would like to note that our protocol offers a non-trivial improvement over the protocol offered by Kiayias et al.~\cite{stego1} as in their protocol, they need $O(n)$ secret bits regardless of the security achieved. 

Also, when we elect to make use of the public randomness for the $d$ bits for the extractor seed, we obtain constant overhead as well. In particular, the length of the shared secret key is equal to the length of the message, $n$ bits while attaining $2^{-n}$ security.

In this context of making the seed of the extractor public, we would like to explain our model and clarify the implications of making the seed public. In our model for steganography, we assume that the communication channel is not adversarially controlled. In particular, the adversary is not allowed to reconfigure the channel distributions once the seed has been made public. In this sense, the channel is chosen and fixed first, then a seed $s$ is chosen uniformly at random and made public. In other words, we require that the randomness in the seed $s$ is independent of the channel. Indeed, in a stronger model where the adversary does have the ability to readapt the channel distributions, we would need to keep the seed private. From our above discussion, we can see that our stegosytem of Section~\ref{sec:stegosystem} is still $(\epsilon,\delta)$-correct and $(\epsilon,\delta)$-secure when the seed $s$ is public.

\begin{theorem}
For any $\epsilon,\delta>0$, message $\vec{m}\in\left\{ 0,1\right\}^{n}$ consider the stegosystem $(SK,SE,SD)$ of Section~\ref{sec:stegosystem} under the  parameter constraints $\epsilon_{ext} \leq \left(\frac{\epsilon}{6 \ell^2}\right)^3$ and $\rho \geq 2n^{c}\log(3 \ell \epsilon^{-1})$ for some constant $c>1$. Then for every channel, if the key $\kappa^{\text{otp}} \in_R \{0,1\}^n$ is private and the seed $s \in_R \{0,1\}^n$ is public, then it holds that the stegosystem is $(\epsilon,\delta)$-correct and $(\epsilon,\delta)$-secure. Here, $\epsilon_{ext}$ is the extractor error and $\ell=\lceil n/c\log n\rceil$ for some constant $c>1$. The stegosystem exhibits $O(1)$ overhead, the length of the shared private key is equal to the length of the message. 
\end{theorem}

\section{A provably secure stegosystem for longer messages}

In Appendix~\ref{appendix_longermessages} we show how to apply the ``one-time'' stegosystem 
of Section~\ref{sec:stegosystem} together with  a pseudorandom 
generator so that longer messages can be transmitted as shown by~\citet{stego1}.

{\small
\bibliographystyle{plainnat} %For the package natbib
\bibliography{s-extractors}}

\appendix

\section{Security Proof}
\label{appendix_secproof}

In this section, we provide the proof for Theorem~\ref{thm:security} from Section~\ref{sec:security}.

\begin{theorem}
For any $\epsilon,\delta>0$, message $\vec{m}\in\left\{ 0,1\right\}^{n}$ consider the stegosystem $(SK,SE,SD)$ of Section~\ref{sec:stegosystem} under the  parameter constraint  $\epsilon_{ext} \leq \left(\frac{\epsilon}{3 \ell}\right)^3$.
Then it holds that the stegosystem is $(\epsilon,\delta)$-secure where $\epsilon_{ext}$ is the extractor error and $\ell=\lceil n/c\log n\rceil$ for some constant $c>1$.
\end{theorem}

\begin{proof} 
We start the encoding procedure $SE$ with history $h$ which embeds 
message blocks into the channel using rejection sampling. We want to show that the statistical distance between the output of $SE$ and $C_{h}^{\lambda}$ is given by 
$$
\Delta\left[SE(\kappa,\vec{m};\mathcal{O}),C_{h}^{\lambda}\right] \leq \epsilon
$$
where $\lambda$ is the length of the output by procedure $SE$. 

First, we define some notation to capture the operation of the procedure $SE$.
Let $C_{{1}}$ denote the distribution at depth
$1$ that results by sampling $\vec{c_{1}}\gets C_{h}^{\pi}$; $C_{2}$
denotes the distribution at depth $2$ that
results by sampling $\vec{c_{1}}\gets C_{h}^{\pi}$ and $\vec{c_{2}}\gets C_{h\circ\vec{c_{1}}}^{\pi}$.
We likewise define $C_{\tau}$ for $\tau \leq \ell$. We define the
random variables $R_1, \cdots, R_\tau$ obtained by rejection sampling
in the same fashion. To be precise, for a message
$\vec{m}^{\prime}=\kappa^{\text{otp}}\oplus\vec{m}=\vec{m_{1}}^{\prime}\circ\vec{m_{2}}^{\prime}\circ\cdots\circ\vec{m_\ell}^{\prime}$
and $|\vec{m_\tau}^{\prime}| = c \log n$ we define
\[
{C}_{1} \triangleq  C_{h}^{\pi}\,, \quad {C}_{\tau} \triangleq  C_{h\circ C_1 \circ
  \cdots \circ C_{\tau-1}}^{\pi}\,,
\]
for $\tau \in \{2, \ldots, \ell\}$. Likewise, we define the random variables
  $R_\tau$:
\[
R_{1} \triangleq
\rejsam_{h}^{E_{s}(\cdot)}\left(\vec{m_1}^{\prime},\rho\right)\,, \quad
R_{\tau} \triangleq \rejsam_{h\circ {R_{1}}\circ\cdots\circ {R_{\tau-1}}}^{E_{s}(\cdot)}\left(\vec{m_{\tau}}^{\prime},\rho\right)\enskip.
\]
Finally, in anticipation of the proof below, we define a ``hybrid''
random variable 
$$
H_\tau = \rejsam_{h\circ {C_{1}}\circ\cdots\circ
  {C_{\tau-1}}}^{E_{s}(\cdot)}\left(\vec{m_{\tau}}^{\prime},\rho\right)
$$
which corresponds to the distribution obtained by selecting $C_1,
\ldots, C_{\tau-1}$ from the natural channel distribution, and then
selecting the $\tau$th channel element via rejection sampling.

Now, let us analyze the implications of picking a uniformly random seed $s\in_{R}\left\{ 0,1\right\}^{d}$ for the extractor as we do in our construction. Recall that $\epsilon_{ext}$ denotes the error of the extractor. First, we show that
for each depth $\tau$, the probability mass of distributions for which the extractor coupled with the seed $s$ yields a $\sqrt[3]{\epsilon_{ext}}$-biased function is large. 

We say that a channel distribution
$C$ is $\left(s,\sqrt[3]{\epsilon_{ext}}\right)\textrm{-good}$
if $E_{s}$ is $\sqrt[3]{\epsilon_{ext}}$-biased on $C$. Otherwise we say that the distribution
$C$ is $\left(s,\sqrt[3]{\epsilon_{ext}}\right)$-bad. With this
definition in place, recall that a strong extractor has the property that
for any distribution $C$ on the right domain with sufficient
min-entropy,
\begin{equation}
\label{eq:single-good}
\Pr_{s}[\text{$C$ is $\left(s,\sqrt[3]{\epsilon_{ext}}\right)$ -bad}]\leq
{\epsilon_{ext}^{2/3}}\,.
\end{equation}
Define now the following sets for $\tau \in \{0,\cdots,\ell-1\}$:
$$
G_{s}^{\tau}=\left\{ \left(\vec{c_{1}},\vec{c_{2}},\cdots,\vec{c_{\tau}}\right)\mid {C_{h\circ\vec{c_{1}}\circ\vec{c_{2}}\circ\cdots\circ\vec{c_{\tau}}}^\pi}\textrm{ is }\left(s,\sqrt[3]{\epsilon_{ext}}\right)\textrm{-good}\right\} 
$$
and 
$$
B_{s}^{\tau}=\left\{ \left(\vec{c_{1}},\vec{c_{2}},\cdots,\vec{c_{\tau}}\right)\mid {C_{h\circ\vec{c_{1}}\circ\vec{c_{2}}\circ\cdots\circ\vec{c_{\tau}}}^\pi}\textrm{ is }\left(s,\sqrt[3]{\epsilon_{ext}}\right)\textrm{-bad}\right\}, 
$$
where $|\vec{c_{i}}| = {\pi}$. The two sets $G_{s}^{\tau}$ and $B_{s}^{\tau}$ denote the collection of $\left(s,\sqrt[3]{\epsilon_{ext}}\right)\textrm{-good}$
and $\left(s,\sqrt[3]{\epsilon_{ext}}\right)\textrm{-bad}$ distributions at depth
$\tau$, respectively.
Let $\mu\left(B_{s}^{\tau}\right)$ denote $\Pr\left[C_{h}^{\tau\pi} \in
  B_{s}^{\tau} \right]$, the total probability mass of the set $B_{s}^{\tau}$. Define
$\mu\left(G_{s}^{\tau}\right)$ similarly. Observe that in light of
Equation~\ref{eq:single-good} above, the expected mass of $B_{s}^{\tau}$ over the choice
of a uniform seed $s$ is
$\mathbb{E}_{s}\left[\mu\left(B_{s}^{\tau}\right)\right]\leq {\epsilon_{ext}^{2/3}}$.
By Markov's inequality 
$\Pr_{s}\left[\mu\left(B_{s}^{\tau}\right)\geq\sqrt[3]{\epsilon_{ext}}\right]\leq\sqrt[3]{\epsilon_{ext}}$
and, then, by the union bound we conclude
\[
\Pr_{s}\left[ \exists \tau < \ell \mid \mu\left(B_{s}^{\tau}\right)\geq\sqrt[3]{\epsilon_{ext}} \right] \leq \ell \sqrt[3]{\epsilon_{ext}}\enskip.
\]
where $\ell=\lceil n/c\log n\rceil$, the number of message blocks. We say that
a seed $s$ is \emph{good} if $\forall \tau \in \{1,2,\cdots,\ell\},
\enskip\mu\left(G_{s}^{\tau}\right)\geq1-\sqrt[3]{\epsilon_{ext}}$. To
summarize the discussion above, for randomly
chosen $s$,
\[
\Pr_{s}\left[s \textrm{ is good}\right] \geq 1-\ell\sqrt[3]{\epsilon_{ext}}\enskip.
\]

Now, fix a good seed $s$. We will now prove that for a good
seed $s$,
\begin{equation}\label{eq:induction}
\Delta\left[\left(C_{1},C_{2},\cdots,C_{\ell}\right),\left(R_{1},R_{2},\cdots,R_{\ell}\right)\right]\leq \ell \cdot\left(3\sqrt[3]{\epsilon_{ext}}\right)\enskip.
\end{equation}
We prove this by induction on $\tau$, the number of message
blocks. When $\tau=1$,
$$\Delta\left[C_{1},R_{1}\right]\leq 2\sqrt{\epsilon_{ext}} \leq
2\sqrt[3]{\epsilon_{ext}}\,,$$
as desired. In general, assuming 
\begin{eqnarray*}
\Delta\left[\left(C_{1},C_{2},\cdots,C_{\tau}\right),\left(R_{1},R_{2},\cdots,R_{\tau}\right)\right] & \leq & \tau\cdot\left(2\sqrt[3]{\epsilon_{ext}}\right)\enskip.
\end{eqnarray*}
for a particular value $\tau$, we wish to establish the inequality for
$\tau+1$. Observe that
\begin{align*}
  \Delta&\bigl[\left(C_{1},C_{2},\cdots, C_{\tau+1}\right),\left(R_{1},R_{2},\cdots,R_{\tau+1}\right)\bigr]  & \\
  &\leq
  \Delta\left[\left(C_{1},C_{2},\cdots,C_{\tau}\right),\left(R_{1},R_{2},\cdots,R_{\tau}\right)\right]
  + \;\Exp_{C_1, \ldots,  C_{\tau}}\bigl[\Delta\left[C_{\tau+1},H_{\tau+1}\right]\bigr] \textrm{(Lemma~\ref{lemma1})}&\\
  &\leq \tau\cdot\left(2\sqrt[3]{\epsilon_{ext}}\right)+\;\Exp_{C_1, \ldots,
   C_{\tau}}\bigl[\Delta\left[C_{\tau+1},H_{\tau+1}\right]\bigr]\quad\text{(by induction.)}
 &
\end{align*}
As for the expectation $\Exp_{C_1, \ldots,
  C_{\tau}}\bigl[\Delta\left[C_{\tau+1},H_{\tau+1}\right]\bigr]$,
observe that \\~\\
$\Exp_{C_1, \ldots,
  C_{\tau}}\bigl[\Delta\left[C_{\tau+1},H_{\tau+1}\right]\bigr]$
\begin{align*}
{} & \leq
\Pr[(C_1, \ldots, C_\tau) \in G^\tau_s] \cdot
\Exp\left[\Delta\left[C_{\tau+1},H_{\tau+1}\right] \mid (C_1, \ldots, C_\tau) \in
  G^\tau_s\right]\\
&\phantom{\leq} + \Pr[(C_1, \ldots, C_\tau) \in B^\tau_s]\cdot \Exp[\Delta\left[C_{\tau+1},H_{\tau+1}] \mid (C_1, \ldots, C_\tau) \in
  G^\tau_s\right]\\
& \leq \Exp[\Delta\left[C_{\tau+1},H_{\tau+1}] \mid  (C_1, \ldots, C_\tau) \in
  G^\tau_s\right] + \Pr[(C_1, \ldots, C_\tau) \in B^\tau_s]\\
& \leq \sqrt[3]{\epsilon_{ext}} + \sqrt[3]{\epsilon_{ext}}\,,
\end{align*}
as $s$ is good. We can conclude that for a good seed $s$, 
\[
\Delta\left[\left(C_{1},C_{2},\cdots,C_{\tau}\right),\left(R_{1},R_{2},\cdots,R_{\tau}\right)\right]\leq \tau \cdot\left(2\sqrt[3]{\epsilon_{ext}}\right),
\]
for any $\tau \leq \ell$. The total statistical distance is now given by
\begin{align*}
\Delta&\left[\left(C_{1},C_{2},\cdots,C_{\ell}\right),\left(R_{1},R_{2},\cdots,R_{\ell}\right)\right]\\
& = \Delta\left[\left(C_{1},C_{2},\cdots,C_{\ell}\right),\left(R_{1},R_{2},\cdots,R_{\ell}\right)\right]\mid_{\textrm{$s$ \emph{good}}}\cdot\Pr[\text{$s$ \emph{good}}]+\\
 &  \quad \;\Delta\left[\left(C_{1},C_{2},\cdots,C_{\ell}\right),\left(R_{1},R_{2},\cdots,R_{\ell}\right)\right]\mid_{\text{$s$ not \emph{good}}}\cdot\Pr[\text{$s$ not \emph{good}}]\\
 & \leq \ell\cdot\left(2\sqrt[3]{\epsilon_{ext}}\right)\cdot
 1+1\cdot(\ell\sqrt[3]{\epsilon_{ext}}) \leq 3 \ell \sqrt[3]{\epsilon_{ext}}  \leq \epsilon\,.
 \end{align*}
The last inequality is because of the fact that $\epsilon_{ext} \leq \left(\frac{\epsilon}{3 \ell}\right)^3$. Thus,
$$\Delta\left[SE(\kappa,\vec{m};\mathcal{O}),C_{h}^{\lambda}\right] \leq \epsilon$$
and the theorem follows by the definition of insecurity.
\end{proof}
\BBB
\section{A provably secure stegosystem for longer messages}
\label{appendix_longermessages}

In this section we show how to apply the ``one-time'' stegosystem 
of Section~\ref{sec:stegosystem} together with  a pseudorandom 
generator so that longer messages can be transmitted as shown by~\citet{stego1}.
\begin{definition}
Let $U_k$ denote the uniform distribution over $\{0,1\}^k$.
A polynomial time deterministic algorithm $G$ is a pseudorandom
generator (PRG) if the following conditions are satisfied:
\begin{description}
\item[Variable output] For all seeds $x \in \{0,1\}^\ast$ and $y \in
\mathbb{N}$, $|G(x,1^y)|=y$.
\item[Pseudorandomness] For every polynomial $p$ the
set of random variables \\ $\{G(U_k, 1^{p(k)} )\}_{k \in {\rm N}}$ is
computationally indistinguishable from the uniform distribution
$\{U_{p(k)}\}_{k \in {\rm N}}$.
\end{description}
\end{definition}

For a PRG $G$ and $0<k<k^{\prime}$, if $A$ is some
statistical test, we define the advantage of $A$ over the PRG as
follows:
\[\mbox{\bf Adv}_{G}^A(k,k^{\prime}) = \left| \Pr_{w \gets G(U_k, 1^{k^{\prime}})}
[A(w) = 1] - \Pr_{w \gets U_{k^{\prime}}}[A(w) = 1]\right|.
\] 
The insecurity of the above PRG $G$ against all statistical tests $A$ computable by circuits of size $\leq P$ is then defined as
$$
\mbox{\bf{InSec}}_{G}(k,k^{\prime};P) =
\max_{A \in \mathcal{A}_P}\{\mbox{\bf{Adv}}_{G}^{A} (k,k^{\prime})\}
$$
where $\mathcal{A}_P$ is the collection of statistical tests computable by circuits of size $\leq P$. 

It is convenient for our application that typical PRGs have a procedure $G'$ such that if $z = G(x, 1^y)$, it holds
that $G(x, 1^{y+y'}) = G'(x, z, 1^{y'})$ (i.e., if one maintains $z$,
one can extract the $y'$ bits that follow the first $y$ bits without
starting from the beginning). 

Consider now the following stegosystem $S' = (SK', SE', SD')$ that can be
used for steganographic transmission of longer messages using the one-time stegosystem $S = (SK, SE, SD)$ as defined in Section~\ref{sec:stegosystem}. $S'$ can handle messages of length polynomial in the security parameter $k$ and employs a PRG $G$. 
The two players Alice and Bob, share a key of length $k$ denoted by
$x$. The function $SE'$ is given input $x$ and the message $m \in\{0,1\}^{\nu}$ to be transmitted of length $\nu = p(k)$ for some fixed polynomial $p$. $SE'$ in turn employs the PRG $G$ to extract $k^{\prime}$ bits (it computes $\kappa = G(x, 1^{k^{\prime}})$, $|\kappa|=k^{\prime}$). The length $k^{\prime}$
is selected to match the number of key bits that are required to
transmit the message $m$ using the one-time stegosystem of
Section~\ref{sec:stegosystem}.  Once the key $\kappa$ of length $k^{\prime}$ is produced by the
PRG, the procedure $SE'$ invokes the one-time stegosystem on input $\kappa,
m, h$. The function $SD'$ is defined in a straightforward way based on $SD$.

The computational insecurity of the stegosystem $S^{\prime}$ is defined by adapting the definition of information theoretic stegosystem security from Section \ref{sec:stegodef} for the computationally bounded adversary as follows:
$$
\mbox{\bf{InSec}}_{S^{\prime}}(k,k^{\prime};P) =  \max_{\mathcal{A} \in \mathcal{A}_P }\{\mbox{\bf{Adv}}_{S'}^\mathcal{A}(k,k^{\prime})\}\,,
$$
this maximum taken over all adversaries $\mathcal{A}$, where $SA_1$ and $SA_2$ have circuit size $\leq P$ and the definition of advantage $\mbox{\bf{Adv}}_{S'}^\mathcal{A}(k,k^{\prime})$ is obtained by suitably modifying the definition of $\mbox{{\bf Adv}}_S^\mathcal{A}(k)$ in Section~\ref{sec:stegodef}. In particular, we define a new adversarial game  $G^\mathcal{A}(1^{k},1^{k^{\prime}})$ which proceeds as the previous game $G^\mathcal{A}(1^k)$ in Section~\ref{sec:stegodef} except that in this new game $G^\mathcal{A}(1^{k},1^{k^{\prime}})$, algorithms $SA_1$ and $SA_2$ receive as input the security parameter $k^{\prime}$ and $SE^{\prime}$ invokes $SE$ as $SE(\kappa,m^\ast; \mathcal{O})$ where $\kappa = G(x, 1^{k^{\prime}})$.

\begin{theorem}
  The stegosystem $S' = (SK', SE' , SD')$ is provably secure in the model
  of \cite{DBLP:conf/crypto/HopperLA02} (steganographically secret against chosen hiddentext attacks); in particular employing a PRG $G$ to transmit a message $m$ we get
$\mathbf{InSec}_{S'}(k,k^{\prime};P) \leq
  \mathbf{InSec}_{G}(k,k^{\prime};P)+\mathbf{InSec}_{S'}(k^{\prime})$ 
where $\mathbf{InSec}_{S'}(k^{\prime})$ is the information theoretic insecurity defined in Section \ref{sec:stegodef} and $|m|=\ell(k^{\prime})$.
\end{theorem}

\section{Omitted proofs}
\label{appendix_omittedproof}

\begin{proof}[\autoref{lemma1}]
%It is easy to observe that the lemma holds if $X, X^{\prime}$ and $Y, Y^{\prime}$ are independent. 
%Let us consider the situation where $Y$ is dependent on $X$ and $Y^{\prime}$ is dependent on $X^{\prime}$. 
For $x \in \mathcal{X}$ denote $\Pr[X=x]$ by $P_{x}$ and $\Pr[Y_x =
y]$ by $P_{y|x}$. Define $P_{x}^{\prime}$ and $P_{y|x}^{\prime}$
similarly. Then we may compute
\begin{eqnarray*} 
\Delta\left[\left(X,Y\right),\left(X^{\prime},Y^{\prime}\right)\right] & = & \frac{1}{2} \sum_{x \in \mathcal{X},y \in \mathcal{Y}}\left|P_{x}\cdot P_{y|x}-P_{x}^{\prime}\cdot P_{y|x}^{\prime}\right|\\
 & \leq & \frac{1}{2}\sum_{x,y}\left|P_{x}\cdot P_{y|x}-P_{x}\cdot
   P_{y|x}'\right| +\frac{1}{2}\sum_{x,y}\left|P_{x}\cdot P_{y|x}'-P_{x}'\cdot P_{y|x}^{\prime}\right| \\
 & = & \frac{1}{2} \sum_{x,y}P_x \cdot
 \left|P_{y|x}-P_{y|x}^{\prime}\right|+\frac{1}{2}\sum_{x,y}P_{y|x}^{\prime}\cdot\left|P_{x}-P_{x}^{\prime}\right|\\
&=& \Exp_{X} \bigl[\Delta\left[Y_{X},Y^{\prime}_{X}\right]\bigr]
 + \Delta\left[X,X^{\prime}\right]\,.
\end{eqnarray*}
%where the triangle inequality is used in the second line.
\end{proof}
\end{document}